\documentclass[aps,prb,twocolumn,showpacs,groupedaddress]{revtex4}
\usepackage{graphics,epsfig}
\begin{document}
\title{What happens to geometric phase when 
spin-orbit interactions lift band degeneracy?}
\author{Philip B. Allen}
\affiliation{Department of Physics and Astronomy,
             Stony Brook University, Stony Brook, NY
             11794-3800
	    }
\date{\today}
\begin{abstract}
Spin-orbit interaction lifts accidental band degeneracy.  The geometric
phase $\gamma(C)=\pm\pi$ for circuits surrounding a line of such degeneracy
cannot survive completely unchanged.  The change depends on how the spin
is fixed during adiabatic evolution.  For spin fixed along the internal
spin-orbit field, $\gamma(C)$ decreases to zero as the circuit collapses
around the line of lifted degeneracy.  For spin fixed along a perpendicular
axis, the conical intersection persists and $\gamma(C)=\pm\pi$ is unchanged. 
\end{abstract}
\maketitle

Geometric (or Berry) phase \cite{Berry} has become a powerful tool for 
analysis of waves in periodic systems, especially
electrons in crystals \cite{Zak,Vanderbilt,Resta,Sundaram,Haldane,Mikitik}. 
Wavevector $\vec{k}$ provides a space in which adiabatic evolution of
wavefunctions $\psi_n(\vec{k},\vec{r})$ can be studied.  Singular behavior
occurs at band degeneracies where energies $\epsilon_1(\vec{k})=
\epsilon_2(\vec{k})$ are equal.  In crystals with inversion symmetry,
ignoring spin-orbit interactions, degeneracies occur along closed
lines in $\vec{k}$-space \cite{Blount}.  
The periodic part $u_n(\vec{k},\vec{r})=\exp(-i\vec{k}\cdot\vec{r})
\psi_n(\vec{k},\vec{r})$ of $\psi$ is an eigenstate
of ${\cal H}(\vec{k})=(\vec{p}+\hbar\vec{k})^2/2m +V(\vec{r})$.
Let the wavevector $\vec{k}(t)$ be given a time evolution which takes
it on the circuit {\it C}, with $\vec{k}(T)=\vec{k}(0)$.
Now suppose that wavefunction evolution
is determined by the time-dependent Schr\"odinger equation with the
time-dependent Hamiltonian ${\cal H}(\vec{k}(t))$.
The time-evolution is assumed adiabatic, namely
$u_n(\vec{k},\vec{r},t) \propto u_n(\vec{k}(t),\vec{r})$.
Berry's argument shows that $u_n(\vec{k},\vec{r},T)$ differs from
$u_n(\vec{k},\vec{r},0)$ by the factor $\exp[i\gamma(C,T)]$, where
the phase $\gamma(C,T)$ has two parts, 
$\gamma(C)+\gamma(T)$.  The dynamical part
$\gamma(T) = -\int_0^T dt \epsilon_n(\vec{k}(t))/\hbar$ depends on the
time elapsed, and the geometric part
\begin{equation}
\gamma(C)=i\oint_C d\vec{k}\cdot\int d\vec{r} u_n^{\ast}\vec{\nabla}_k u_n
\label{eq:geomphase}
\end{equation}
is invariant and intrinsic to the circuit and the band properties.
In particular, $\gamma(C)=\pm\pi$ if {\it C} encloses one (or an odd
number) of degeneracy lines.  This change of wavefunction sign is
familiar from other problems where a circuit of adiabatic evolution
surrounds a conical intersection.  Direct evaluation of 
Eq.(\ref{eq:geomphase}) is problematic.  Wavefunctions must be
continuous and single-valued.

Although gauge invariance is not evident in Eq.(\ref{eq:geomphase}),
Berry gave also an alternate form, for a 3-dimensional parameter
space $\vec{k}$, as the flux through a surface {\it S} (bounded by {\it C})
of a vector $\vec{V}_n$.  
\begin{equation}
\gamma(C)=-\int_S d\vec{S}_{\vec{k}}\cdot\vec{V}_n
\label{eq:Sint}
\end{equation}
\begin{equation}
\vec{V}_n = {\rm Im}\sum_m \frac{\langle n|\vec{\nabla}_{\vec{k}}{\cal H}
|m\rangle \times \langle m|\vec{\nabla}_{\vec{k}}{\cal H} |n\rangle}
{[\epsilon_m(\vec{k}) -\epsilon_n(\vec{k})]^2}
\label{eq:V}
\end{equation}
The gauge invariance of this vector is evident.  
Conditions of continuity and single-valuedness of
wavefunctions are no longer required.
If the circuit surrounds a singularity described
by a $2\times2$ effective Hamiltonian, then
the flux equals half the solid angle $\Omega(C)$ subtended in
an appropriate scaled space by the circuit as seen from the
point of singularity.  The appropriate scaled space is
the one in which the $2 \times 2$ Hamiltonian for states
near the conical intersection has the form ${\cal H}_{\rm eff}=\vec{R}
\cdot\vec{\sigma}$ in terms of scaled coordinates
$\vec{R}=(X,Y,Z)$ and Pauli matrices $\vec{\sigma}=
(\sigma_x,\sigma_y,\sigma_z)$.  This method will be used
twice in this paper.  The eigenvalues
are $\pm R=\rho_z R$, where the quantum number $\rho_z=\pm 1$
is introduced as a branch index.  The geometric phase
is then $\gamma(C)=-\rho_z \Omega(C)/2$.

Mikitik and Sharlai \cite{Mikitik}
provide convincing evidence that the geometric phase $\pm\pi$
is seen experimentally as a shift in the
semiclassical quantization condition \cite{Kosevich}
determining the de Haas-van Alphen oscillations.
An extreme experimental case is the shifted quantum Hall oscillations
originating from orbits near the ``Dirac points'' 
in graphene \cite{Graphene1,Graphene2}.  
The shifts of quantization condition 
occur for electron orbits (in a $\vec{B}$-field)
which surround a degeneracy line (or point, for graphene.)
They also argue \cite{Mikitik2} that spin-orbit
effects can mostly be ignored.  This is correct for lighter
elements with spin-orbit strength $\xi/\Delta \ll 1$,
$\Delta$ being any other relevant electron scale such as a band gap.
However, the mathematics and the corrections need elucidation.
Spin-orbit coupling destroys band degeneracy lines.  It
is not evident what happens to the geometric phase of $\pm\pi$.

To see the effect of spin-orbit interactions, add to ${\cal H}(\vec{k})$
the piece ${\cal H}_{SO}=(\vec{\sigma}/4m^2c^2)\cdot\vec{\nabla}V \times
(\vec{p}+\hbar\vec{k})$.  Choose some point $\vec{k}^{\ast}$ of
accidental degeneracy, and find energies and eigenstates at nearby
$\vec{k}$-points using degenerate $\vec{k}\cdot\vec{p}$ perturbation theory.
For notational simplicity, $\vec{k}^{\ast}$ is 
the temporary origin of $\vec{k}$.  
The degenerate basis functions $|1\rangle$ and $|2\rangle$ are the
periodic parts $u_1$ and $u_2$ at $\vec{k}=\vec{k}^{\ast}$.  
A phase convention is needed; the coefficients $C_G$ of
the expansion $u(\vec{r})=\sum C_G \exp(i\vec{G}\cdot\vec{r})$
are chosen real.  This requires inversion symmetry, which
is hereafter assumed.
Each state has two spin orientations, so the effective 
Hamiltonian matrix is $4\times 4$, with the form
\begin{equation}
{\cal H}_{\rm eff}=\left(\begin{array}{cc}
\hbar\vec{k}\cdot\vec{v}_a\hat{1}  
& \hbar\vec{k}\cdot\vec{v}_b\hat{1} -i\vec{\xi}\cdot\vec{\sigma} \\
\hbar\vec{k}\cdot\vec{v}_b\hat{1} +i\vec{\xi}\cdot\vec{\sigma} 
& -\hbar\vec{k}\cdot\vec{v}_a\hat{1} \end{array}\right)
\label{eq:Heff}
\end{equation}
where $\hat{1}$ and $\vec{\sigma}$ are $2\times 2$ matrices in spin space.
Terms proportional to the $4\times 4$ unit matrix do not mix or split the
states and are omitted.   The vector $\vec{v}_a$ is $(\vec{v}_1 - \vec{v}_2)/2$,
where $\vec{v}_n$ is the band velocity $\vec{\nabla}_k \epsilon_n/\hbar$
at the degeneracy $\vec{k}^{\ast}$.  The vector $\vec{v}_b$ is the off-diagonal
term $\langle 2|\vec{p}/m|1\rangle$, which is pure real 
since $C_G$ is real.  The vector $i\vec{\xi}$ is $\langle2|\vec{\nabla}V
\times\vec{p}|1\rangle/4m^2 c^2$.  
This is pure imaginary since there is also
time-reversal symmetry, under an assumption of
no magnetic order or external $\vec{B}$-field.  Thus 
three real vectors, $\vec{v}_a$, $\vec{v}_b$, and $\vec{\xi}$, determine
the bands near $\vec{k}^{\ast}$.  
The vector $\vec{\xi}$ is a close analog to angular momentum.  
Consider a system with two degenerate 
$p$-states $|x\rangle$ and $|y\rangle$.  The angular momentum
operator $\vec{L}$ has an imaginary off-diagonal element.  The mixed states
$|x\rangle\pm i |y\rangle$
are eigenstates of $\vec{L}$ with 
$\langle\vec{L}\rangle=\pm m\hbar\hat{z}$.
The magnitude $m$ deviates from $1$ if the point symmetry is less
than spherical.  The vector $\vec{\xi}$ will be called the ``orbit moment.''

\begin{figure}
\centerline{\scalebox{0.36}[0.36]{\includegraphics{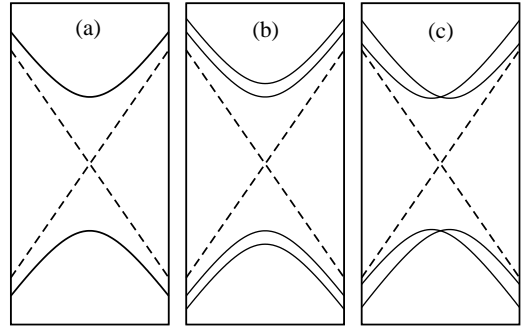}}}
\caption{Energy {\it versus} $|\vec{k}|$ near the degeneracy point,
for (a) no magnetic field, (b) field parallel to $\vec{\xi}$, and
(c) field perpendicular to $\vec{\xi}$.  The dashed lines are for $\xi=0$ 
and $b=0$; solid lines in panel (a) are $\pm\lambda$, which becomes
$\pm\xi$ at the degeneracy point $\vec{k}=0$.
}
\label{fig:1}
\end{figure}

First suppose that $\vec{\xi}=0$.  Since $\vec{v}_a$ and $\vec{v}_b$
are not generally co-linear, they define a direction of $\vec{k}$, namely
$\vec{v}_a \times \vec{v}_b$, along which ${\cal H}_{\rm eff}=0$.  This is
the direction of the line of degeneracy.  After allowing $\vec{\xi}\ne 0$,
 eigenvalues of Eq.(\ref{eq:Heff}) are $\pm\lambda$ where
\begin{equation}
\lambda = \sqrt{\kappa_a^2 + \kappa_b^2 + \xi^2}
\label{eq:lambda}
\end{equation}
with $\kappa_a=\hbar\vec{k}\cdot\vec{v}_a$,
$\kappa_b=\hbar\vec{k}\cdot\vec{v}_b$, and $\xi=|\vec{\xi}|$.
Each eigenvalue belongs to a Kramers doublet of two opposite spin states.
The original degeneracy (without spin-orbit interaction) of 2
(neglecting spin) or 4 (including spin) is lifted everywhere unless
$\vec{\xi}=0$.  This should
happen only at isolated points in the Brillouin zone, not
coinciding with degeneracy lines $\vec{k}^{\ast}$.  No accidental
degeneracies remain, but Kramers degeneracy occurs everywhere.  
Bands near $\vec{k}^{\ast}$ are shown in Fig.\ref{fig:1}.

The geometric phase under consideration involves a circuit $C(\vec{k})$
surrounding the $\vec{k}^{\ast}$ line.  A circular path
in two-dimensional $(\kappa_a,\kappa_b)$-space,
namely $C=(\kappa\cos\phi,\kappa\sin\phi), \ 0\rightarrow\phi
\rightarrow 2\pi$ is the simplest realization.  
To calculate $\gamma(C)$, separate Eq.(\ref{eq:Heff})
into two similar $2\times2$ submatrices by choosing basis states with
spins polarized along $\vec{\xi}$, which will be 
used as the $z$-axis of spin space.  The submatrices are
\begin{equation}
{\cal H}_{{\rm eff}}^{\pm}=\left(\begin{array}{cc}
\kappa_a & \kappa_b \mp i\xi \\
\kappa_b \pm i\xi & -\kappa_a \end{array}\right),
\label{eq:Heffpm}
\end{equation}
where the upper sign goes with spin up, $\sigma_z=1$.  

The circuit can now be considered as
a path $C(\vec{\lambda})$ in a 3-d $\vec{\lambda}$-space,
where $(\lambda_x,\lambda_y,\lambda_z)=(\kappa_b,\sigma_z\xi,\kappa_a)$.
On this circuit, $\lambda$, $\kappa$, and $\xi$ are all constant.
The effective Hamiltonian has the desired scaled form.
The solid angle is $\sigma_z 2\pi(1-\xi/\lambda)$, so the geometric phase is 
\begin{equation}
\gamma(C)=-(\Lambda_z\sigma_z)\pi(1-\xi/\lambda).
\label{eq:zphase}
\end{equation}
where $\Lambda_z=\pm 1$ is the branch index.
This is one of the two main results of this paper.
It shows how spin-orbit splitting destroys the simple phase
of $\pm\pi$ when the circuit has such a small radius that $\xi\sim\lambda$.  
If spin-orbit interaction is weak, it does not need a large orbit to
have $\xi/\lambda \ll 1$ and approach the full simple phase of $\pm\pi$.

This is not the full story.  The choice to evolve at fixed
$\sigma_z$ was arbitrary.  The states of Kramers
doublets can be mixed by arbitrary unitary transformations.
Evolution of a doublet around a circuit introduces not a simple
geometric phase, but a unitary matrix.  The $\gamma(C)$ phases
just computed are actually the diagonal elements $\exp(\pm i\gamma(C))$
of a $2\times2$ unitary matrix in the representation with spin
quantized along $\vec{\xi}$.  It will emerge below that this is indeed the 
correct adiabatic evolution of the Kramers doublet when
an small magnetic field is imposed along
the $\vec{\xi}$ direction.

Berry's original argument 
assumed that ${\cal H}$ had a discrete spectrum along {\it C}.
There is a physically natural way to
retain this.  Magnetic fields present in experiment
since they are used to cause cyclic evolution
in $\vec{k}$-space.  Magnetic fields also lift Kramers degeneracy.
The simplest theoretical device is to
add to ${\cal H}_{\rm eff}$ a Zeeman 
term ${\cal H}_Z =-\vec{b}\cdot\vec{\sigma}$
coupling only to spin.

To proceed further,
an explicit representation of eigenstates is needed.
Eigenstates of the effective Hamiltonian (\ref{eq:Heffpm}),
labeled by energy $\pm\lambda$ and $\sigma_z=\uparrow,\downarrow$ are
chosen as
\begin{equation}
|s\rangle=|-\lambda,\uparrow\rangle=\frac{1}{n}\left (\begin{array}{c}
-\kappa_b+i\xi \\ \kappa_a+\lambda \end{array} \right)
\otimes|\uparrow\rangle,
\label{eq:lowerup}
\end{equation}
\begin{equation}
|t\rangle=|-\lambda,\downarrow\rangle=\frac{1}{n}\left(\begin{array}{c}
-\kappa_b-i\xi \\ \kappa_a+\lambda\end{array}\right)
\otimes|\downarrow\rangle,
\label{eq:lowerdown}
\end{equation}
\begin{equation}
|u\rangle=|+\lambda,\uparrow\rangle=\frac{1}{n}\left (\begin{array}{c}
\kappa_a+\lambda \\ \kappa_b+i\xi \end{array} \right)
\otimes|\uparrow\rangle,
\label{eq:upperup}
\end{equation}
\begin{equation}
|v\rangle=|+\lambda,\downarrow\rangle=\frac{1}{n}\left(\begin{array}{c}
\kappa_a+\lambda \\ \kappa_b-i\xi\end{array}\right)
\otimes|\downarrow\rangle.
\label{eq:upperdown}
\end{equation}
These are written as direct product of spatial times spin two-vectors. 
The normalization is $n=\sqrt{2\lambda(\lambda+\kappa_a)}$.
As long as $\xi$ is non-zero, $1/n$ is non-singular and
these are smooth, single-valued functions of
$(\kappa_a,\kappa_b)$, unique except for an arbitrary overall phase,
which cannot alter $\gamma(C)$.
The lower Kramers doublet $|s\rangle,|t\rangle$ has ``orbit moments''
$\langle i|\vec{\nabla}V \times\vec{p}/4m^2 c^2|i\rangle
=\mp(\xi/\lambda)\vec{\xi}$
oriented antiparallel to spin, while the
upper Kramers doublet $|u\rangle,|v\rangle$ has
identical orbit moments except oriented parallel to spin.

Now the Zeeman term is added.
Diamagnetic coupling is neglected.  Without loss of generality,
the part of the field $\vec{B}=\vec{b}/\mu_B$ perpendicular to $\xi$ can
be used to define the $x$ direction of spin.
The total Hamiltonian in the basis 
$|s\rangle,|t\rangle,|u\rangle,|v\rangle$ is
\begin{equation}
{\cal H}_{\rm tot}=-\left(\begin{array}{cccc}
\lambda+b_z&\frac{\kappa}{\lambda}b_x e^{i\omega}&0&i\frac{\xi}{\lambda}b_x\\
\frac{\kappa}{\lambda}b_x e^{-i\omega}&\lambda-b_z&-i\frac{\xi}{\lambda}b_x&0\\
0&i\frac{\xi}{\lambda}b_x&-\lambda+b_z&\frac{\kappa}{\lambda}b_x e^{-i\omega}\\
-i\frac{\xi}{\lambda}b_x&0&\frac{\kappa}{\lambda}b_x e^{i\omega}&-\lambda-b_z
\end{array}\right)
\label{eq:Htot}
\end{equation}
The factor $(\kappa/\lambda) \exp(i\omega)=\langle s|\sigma_+|t\rangle$
introduces the new angle $\omega$ 
\begin{equation}
e^{i\omega}=\frac{\lambda}{\kappa}-\frac{\xi(\xi-i\kappa_b)}{\kappa(\lambda+\kappa_a)}.
\label{eq:omega}
\end{equation}
As the circuit {\it C} is followed ($\phi$ going from 0 to $2\pi$,
with $\xi,\kappa,\lambda$ constant),
$\omega$ also evolves from 0 to $2\pi$.

If the field $\vec{b}$ is along $z$, the upper and lower Kramers
doublets are not coupled.  The degeneracy is lifted everywhere,
and adiabatic evolution proceeds smoothly on the resulting non-degenerate
states, yielding the phases $\gamma(C)$ of Eq.(\ref{eq:zphase}).  
The previous discussion was correct.
The result \ref{eq:zphase} can also be obtained directly from
Eq.(\ref{eq:geomphase}) using Eqs.(\ref{eq:lowerup},\ref{eq:lowerdown},
\ref{eq:upperup},\ref{eq:upperdown}).
For fields perpendicular to $z$, there is both intra- and
inter-doublet spin mixing, according to Eq.(\ref{eq:Htot}).
To first order, since $\vec{b}\ll\lambda$,
inter-doublet mixing terms $\pm i\xi b_x /\lambda$ can
be neglected, giving $2\times2$ effective Hamiltonian
matrices, of the form
\begin{equation}
{\cal H}_{\rm eff}(\vec{b}) =\lambda_z \lambda\hat{1} 
- \left( \begin{array}{cc}
b_z & \frac{\kappa}{\lambda}b_x e^{i\lambda_z \omega} \\
\frac{\kappa}{\lambda}b_x e^{-i\lambda_z \omega} & -b_z
\end{array}\right)
\label{eq:Heffb}
\end{equation}
The eigenvalues are
\begin{equation}
\pm \lambda \pm \mu \ \ \ {\rm where} \ \mu^2=b_z^2 + \frac{\kappa^2}
{\lambda^2}b_x^2
\label{eq:beigenvalues}
\end{equation}
These eigenvalues have an interesting feature: 
at the degeneracy point $\kappa=0$, in the center of circuit {\it C},
$\mu=0$ and Kramers degeneracy is {\bf not} lifted,
provided $\vec{b}$ is
perpendicular to $\vec{\xi}$.  The states at $\vec{k}^{\ast}$
have anisotropic $g$ factors which vanish in two directions.
The vanishing Zeeman splitting means 
that a conical intersection, hidden unless $\vec{b}\ne 0$, 
exists exactly where the original band
intersection (for $\xi=0$) was located.  This also yields a simple
geometrical phase of $\pm\pi$.  Bands for $\vec{b}\parallel\vec{\xi}$
and $\vec{b}\perp\vec{\xi}$ are shown in Fig.\ref{fig:1} panels (b)
and (c).

A full calculation of $\gamma(C)$ for the 4 new eigenstates
of Eq.(\ref{eq:Htot}) is difficult.  
The Berry method of solid angle works when the basis
functions $|1\rangle$, $|2\rangle$ of the $2\times2$
effective Hamiltonian are fixed at $\vec{k}^{\ast}$,
whereas the basis functions $|s\rangle,|t\rangle$ or
$|u\rangle,|v\rangle$ used in Eq.(\ref{eq:Heffb}) depend
on $\vec{k}$.  However, the most important limit
remaining to be resolved is when the circuit radius $\kappa$
is small relative to spin-orbit splitting $\xi$.  In this limit,
the basis functions loose their $\vec{k}$-dependence. 
The relevant scaled parameters are $\vec{\mu}=((\kappa/\lambda)
b_x\cos\omega,-\lambda_z(\kappa/\lambda)b_x\sin\omega,b_z)$.
The circuit parameterized by $\phi$ is equally well
parameterized by $\omega$ which evolves from 0 to $2\pi$.
The solid angle in $\vec{\mu}$-space is $2\pi\Lambda_z(1-b_z/\mu)$,
so the geometric phase is
\begin{equation}
\gamma(C)=-\pi\beta_z\Lambda_z\left(1-\frac{b_z}{\sqrt{b_z^2+
(\kappa/\lambda)^2 b_x^2}}\right),
\label{eq:finalphase}
\end{equation}
where $(\Lambda_z,\beta_z)$ are the two branch indices in the
eigenvalue $\pm\lambda\pm\mu=\Lambda_z\lambda+\beta_z\mu$.  
This is the other main result of this paper.
If $b_z=0$, the full
geometric phase $\gamma(C)=\pm\pi$ is restored no matter
how small the circuit radius.  Even though the degeneracy
was lifted by spin-orbit interactions, the hidden conical
intersection exposed by a Zeeman field controls the result.

I thank A. G. Abanov and M. S. Hybertsen for help.
I thank the students of Phy556 who were subjected to preliminary
versions of this work.
This work was supported in part by NSF grant no.
NIRT-0304122. 

\end{document}